\documentclass[10pt,onecolumn]{revtex4-2}
\usepackage{amssymb}
\usepackage{graphicx}
\usepackage{subcaption}
\usepackage{float}
\usepackage{bm}
\pdfoutput=1
\usepackage{epstopdf}
\usepackage[T1]{fontenc}
\usepackage{selinput}
\SelectInputMappings{%
  aacute={‡},
  ntilde={–},
  Euro={Û}
}
\usepackage{babel}

\begin{document}

\title{Materia cu‡ntica y din‡mica por mediciones}
\author{Santiago F. Caballero-Ben\'{\i}tez\footnote{Corresponding author: scaballero@fisica.unam.mx}  
 }

\affiliation{ 
Instituto de F\'{\i}sica, LSCSC-LANMAC, Universidad
Nacional Aut\'onoma de M\'exico, Apdo. Postal 20-364, M\'exico D.
F. 01000, M\'exico. }

\date{\today}
\begin{abstract}
This article discusses some details of the course on "Quantum matter and measurement induced dynamics" given in the Summer School of Physics XXIX at UNAM in 2022. The notes describe useful concepts to study the dynamics induced by photon losses, the method for simulation (quantum trajectories) is summarized and details of models in optical lattices and high-Q cavities are given. The notes are in Spanish.

En este art'culo se discuten algunos detalles del curso sobre "Materia cu‡ntica y din‡mica inducida por medici—n" de la escuela de verano de F'sica XXIX (2022) en la UNAM. Las notas describen conceptos œtiles para estudiar la din‡mica emergente por efectos de medici—n de fotones, se resume el mŽtodo para simulaci—n (trayectorias cu‡nticas) y se dan detalles de modelos de materia cu‡ntica y cavidades de alta reflectancia. 
\end{abstract}

\maketitle
Las notas del curso son una introducci—n sobre el mŽtodo implementado de trayectorias cu‡nticas para estudiar la din‡mica inducida en sistemas en redes —pticas y cavidades de alta reflectancia. Se estudia la din‡mica inducida en el rŽgimen donde la tasa de pŽrdida de fotones ($\kappa$) es grande con respecto a la desinton'a de la cavidad entre el bombeo y la frecuencia de la cavidad  ($\Delta=\omega_c-\omega_p$), $\kappa\gg|\Delta|$:
\begin{itemize}
\item{Materia cu‡ntica, modelos de Hubbard y redes —pticas}
\item{Ingenier'a de estados cu‡nticos con 
correlaciones v'a retro-acci—n de la pŽrdida de
fotones}
\item{Retroalimentaci—n y control de criticalidad}
\item{Conclusiones}
\end{itemize}

\section{Materia cu‡ntica, modelos de Hubbard y redes —pticas}

Los recientes avances en el control de los sistemas at—micos han permitido generar una nueva generaci—n de simuladores cu‡nticos ~\cite{qsim1,qsim2}. Los modelos m‡s simples con interacciones de muchos cuerpos t'picos de materia cu‡ntica ultrafr'a en redes —pticas ~\cite{Lewenstein,Jaksch} son el modelo de Hubbard de fermiones  ($\mathcal{H}^f$):
\begin{equation}
\mathcal{H}^f=-J\sum_{\sigma}\sum_{\langle i , j \rangle }\left(\hat{f}_{i\sigma}^\dagger \hat f_{i\sigma}^{\phantom{\dagger}}+\textrm{H.c.}\right)+U\sum_{i}\hat{n}_{i\uparrow}\hat{n}_{i,\downarrow}
\end{equation}
y modelo de Bose-Hubbard ($\mathcal{H}^b$): 
 \begin{equation}
\mathcal{H}^b=-J\sum_{\langle i, j\rangle}(\hat b^\dagger_i\hat b^{\phantom{\dagger}}_j+\mathrm{H.c.})+\frac{U}{2}\sum_i\hat n_i(\hat n_i-1) 
\end{equation}
Los tŽrminos de $J$ representa la energ'a cinŽtica ($J\sim\mathrm{kHz}$) y los tŽrminos con $U$ la energ'a de interacci—n de muchos cuerpos, que puede ir m‡s all‡ de perturbaciones y puede generar transiciones de fase cu‡nticas~\cite{Sachdev}.
Estos modelos representan el comportamiento de los ‡tomos con diferentes propiedades estad'sticas (bosones/fermiones) en segunda cuantizaci—n que est‡n en el l'mite donde los procesos de baja energ'a dominan. Para lograr estos sistemas se emplean varias etapas de enfriamiento~\cite{cooling} hasta llegar a temperaturas de $10^{-9}$ a $10^{-6}$ K y despuŽs son atrapados por l‡seres que generan una estructura de red (red —ptica) que puede ser controlada de manera externa paramŽtricamentee~\cite{Bloch, Schafer}. Los par‡metros del sistema con controlados por las amplitudes de los l‡seres y las colisiones de baja energ'a, t'picamente de onda-s via la longitud de dispersi—n $a$. Dependiendo de los ‡tomos en el experimento, esta longitud de dispersi—n incluso puede ser controlada de manera externa mediante un campo magnŽtico externo constante y la resonancia de Feshbach que se genera~\cite{Feshbach}, $a=a(B)$. Los par‡metros de los modelos se relacionan de la siguiente manera:
\begin{eqnarray}
J&=&\int\mathrm{dx}^D w(\mathbf{x}-\mathbf{x}_i)^*
\left(
\frac{\nabla^2}{2m}-V_{\mathrm{OL}}(\mathbf{x})
\right)
w(\mathbf{x}-\mathbf{x}_j)\;\;
\\
U&=&\frac{4\pi \hbar a}{m}\int\mathrm{dx}^D|w(\mathbf{x}-\mathbf{x}_i)|^4
\end{eqnarray}
donde $i,j$ son los vecinos m‡s cercanos, $w(\mathbf{x})$ son las funciones de Wannier localizadas~\cite{Lewenstein}, $m$ la masa de los ‡tomos fermiones ($\phantom{a}^6$Li~\cite{FLi},$\phantom{a}^{40}$K~\cite{FK}) o bosones ($\phantom{a}^7$Li~\cite{SLi},$\phantom{a}^{23}$Na~\cite{SNa}, $\phantom{a}^{87}$Rb~\cite{SRb})   y $D$ la cantidad de dimensiones espaciales. La longitud de dispersi—n de onda-s es $a=a_{s,\uparrow\downarrow}$ para el caso de fermiones donde se se toman en cuenta colisiones entre diferentes estados magnŽticos at—micos (debido al  bloqueo de Pauli) y en el caso de bosones se consideran ‡tomos con el mismo estado, $a=a_s$.   Otros detalles relevantes sobre la analog'a con sistemas electr—nicos en la aproximaci—n de enlace fuerte y detalles sobre redes —pticas se pueden encontrar en ~\cite{Lewenstein, CaballeroMBCQED}.
T'picamente para una red hiper-cœbica de dimensi—n $D$ uno tiene para el potencial de la red —ptica, 
\begin{equation}
V_{\mathrm{OL}}(\mathbf{x})=\sum_{\eta=1}^DV_0\sin^2(k_\eta x_\eta)
\end{equation} 
donde $k_\eta$ es el vector de onda de la luz laser incidente formando una onda estacionaria en la direcci—n  $\hat e_\eta$, $x_\eta$ es la coordenada (i. e. $x$, $y$, $z$, $\dots$). Para cada ``dimensi—n'' $\eta$ tenemos vectores unitarios $\hat{e}_\eta$. 

\section{Ingenier'a de estados cu‡nticos con 
correlaciones v'a retro-acci—n de la pŽrdida de
fotones}
\begin{figure}[t!]
\begin{center}
\includegraphics[width=0.6\textwidth]{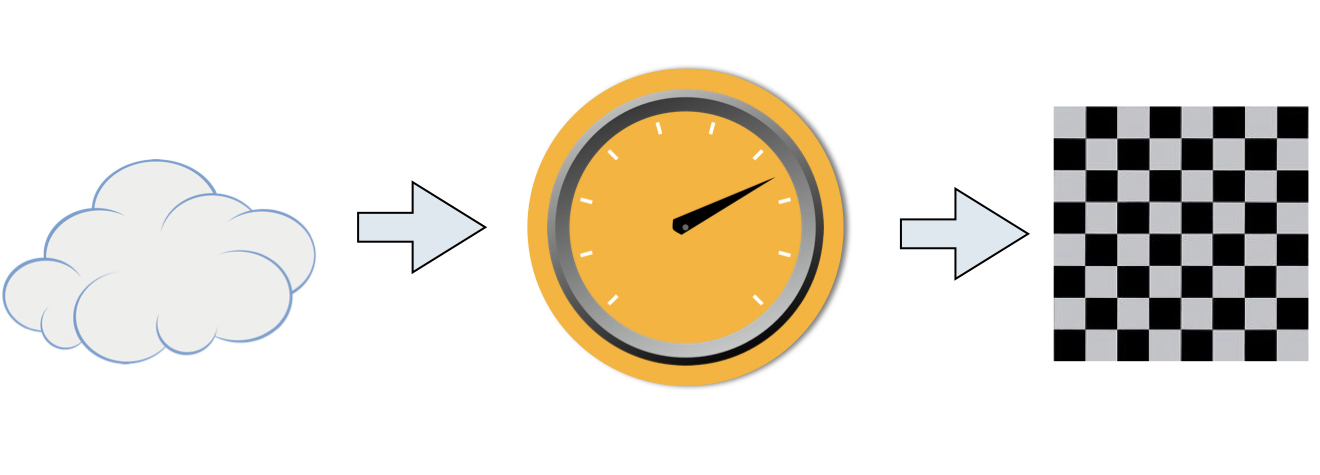}
\end{center}
\caption{Esquema general de  la din‡mica inducida por medici—n. Izquierda: sistema sin estructura. Centro: medici—n dŽbil. Derecha: estructura emergente inducida, i.e. dos modos espaciales en una red cuadrada. }
\label{fig1}
\end{figure}

\subsection{Esquema general de din‡mica por medici—n}
Recientemente se han logrado experimentos de ‡tomos ultrafr'os en redes —pticas cl‡sicas dentro de una cavidad de alta reflectancia~\cite{Hemmerich-LS,Esslinger-LS,Esslinger-LS2}. Estos arreglos t'picamente se han realizado con cavidades de onda estacionaria pero tambiŽn se pueden considerar arreglos con cavidades de ondas viajeras ~\cite{LP9}. TambiŽn existen realizaciones de sistemas sin redes —pticas con mœltiples cavidades ~\cite{Esslinger-CC,Esslinger-CC1} o  cavidades multi-modales~\cite{Kollar}. En principio se puede colocar con posiciones arbitrarias los detectores de fotones que escapan la cavidad. T'picamente los detectores se colocan en los ejes de la cavidad, detr‡s de los espejos  de la cavidad donde la amplitud de la se–al de salida es m‡xima.  En estos sistemas se puede estudiar el l'mite donde la pŽrdida de fotones es alta con respecto a otras escalas de energ'a en el sistema, concretamente la desinton'a $\Delta$. En estas condiciones un escenario a estudiar es si la pŽrdida de fotones que cambia el estado del sistema de manera din‡mica puede inducir estados con estructuras œtiles. Por una estructura œtil nos referimos a un estado que t'picamente el sistema no soporta para bajas energ'as, como puede ser un estado con una simetr'a rota inducida u otro tipo de orden emergente, ver Fig.\ref{fig1}. Para estudiar esta posibilidad se estudia de manera efectiva la evoluci—n temporal condicionada del sistema por la pŽrdida de fotones utilizando el mŽtodo de trayectorias cu‡nticas~\cite{Wiseman-Milburn}. Para esto se construye el siguiente Hamiltoniano no-Hermitiano,
\begin{equation}
\mathcal{H}_{\mathrm{NH}}=\mathcal{H}-i\hat{c}^\dagger\hat{c}
\end{equation}
donde  
\begin{equation}
\hat{c}\approx\frac{\sqrt{2\kappa}g\hat{F}}{\Delta+i\kappa}
\end{equation}
es el operador de salto cu‡ntico de la trayectoria,  $\kappa$ es la tasa de pŽrdida de fotones de la cavidad, $g$ es el acoplamiento luz materia,  $\Delta$ la desinton'a y $\hat {F}$ el operador que codifica la proyecci—n espacial inducida por la luz bombeada al sistema en la materia y la cavidad.  El operador $\hat{F}$ esta dado por la proyecci—n del traslape de los modos bombeados y los modos en la cavidad con componentes espaciales~\cite{QOL-Caballero,QSim-Caballero,Bond-Caballero,Wojciech, Wojciech1,  Wojciech2,Mazzucchi} y considerando adicionalmente grados internos de libertad (i.e. esp'n)~\cite{Mazzucchi1, Mazzucchi2, FQOL-Caballero,SQOL-Caballero},
\begin{equation}
\hat F= \sum_{\sigma,p,c}(\hat D_{\sigma pc}+\hat B_{\sigma pc})
\end{equation}
con modos de densidad,
\begin{equation}
\hat D_{\sigma pc}=\sum_{i}J_{ii}^{\sigma pc}\hat m^\dagger_{\sigma i}\hat m^{\phantom{\dagger}}_{\sigma i}
\end{equation}
y modos de enlace,
\begin{equation}
\hat B_{pc\sigma}=\sum_{\langle i,j\rangle}J_{ij}^{\sigma pc}( \hat m^\dagger_{\sigma i}\hat m^{\phantom{\dagger}}_{\sigma j}+\textrm{H.c.})
\end{equation}
con las constantes de estructura,
\begin{equation}
J_{lm}^{\sigma pc}=\int\mathrm{dx}^Dw(\mathbf{x}-\mathbf{x}_l)^*u_{\sigma,c}(\mathbf{x})^*u_{\sigma,p}(\mathbf{x})w(\mathbf{x}-\mathbf{x}_m)
\end{equation}
La forma particular del operador de salto $\hat{c}$ esta directamente relacionada con la eliminaci—n adiab‡tica de la luz, de tal forma que, la luz se encuentra esclavizada a la materia~\cite{EPJD08, NJPhys2015}.

Adem‡s, se requiere que despuŽs de que un salto cu‡ntico ocurra (la medici—n de un fot—n) el estado del sistema sea renormalizado tal que:
\begin{equation}
|\Psi'\rangle\to\frac{\hat{c}|\Psi\rangle}{\langle\hat c^\dagger\hat{c}\rangle}
\end{equation}
de manera efectiva el Hamiltoniano no Hermitiano se puede escribir como~\cite{Mazzucchi1}:
\begin{equation}
\mathcal{H}_{\mathrm{NH}}\approx\mathcal{H}^{f/b}-i\frac{\kappa g_{\mathrm{eff}}}{\Delta}\left(\hat{F}^\dagger\hat{F}+\hat{F}\hat{F}^\dagger\right)
\end{equation}
con $g_{\mathrm{eff}}=\Delta|g|^2/(\Delta^2+\kappa^2)$ en el l'mite donde $\kappa\gg\Delta$. El nœmero de fotones que escapa de la cavidad es $N_{ph}=\langle\hat{a}^\dagger\hat{a}\rangle\propto\langle\hat{c}^\dagger\hat{c}\rangle$.

Para estudiar los efectos de la din‡mica inducida por medici—n se implementa el siguiente esquema de simulaci—n.
\subsection{Esquema del mŽtodo de trayectorias}
\begin{enumerate}
\item{Iniciar el sistema en el estado base u otro estado arbitrario.}
\item{Generar un intervalo $t$ y un nœmero aleatorio $r$ en el intervalo [0,1]}
\item{Propagar el sistema con el Hamiltoniano no Hermitiano hasta que la norma del estado es igual a $r$.}
\item{Aplicar el operador de salto cu‡ntico (pŽrdida de un fot—n) y renormalizar el estado  con respecto a    
repetir 2 a 4 para un nuevo intervalo $tÕ$} 
\end{enumerate}

\begin{figure}[t!]
\begin{center}
\includegraphics[width=0.6\textwidth]{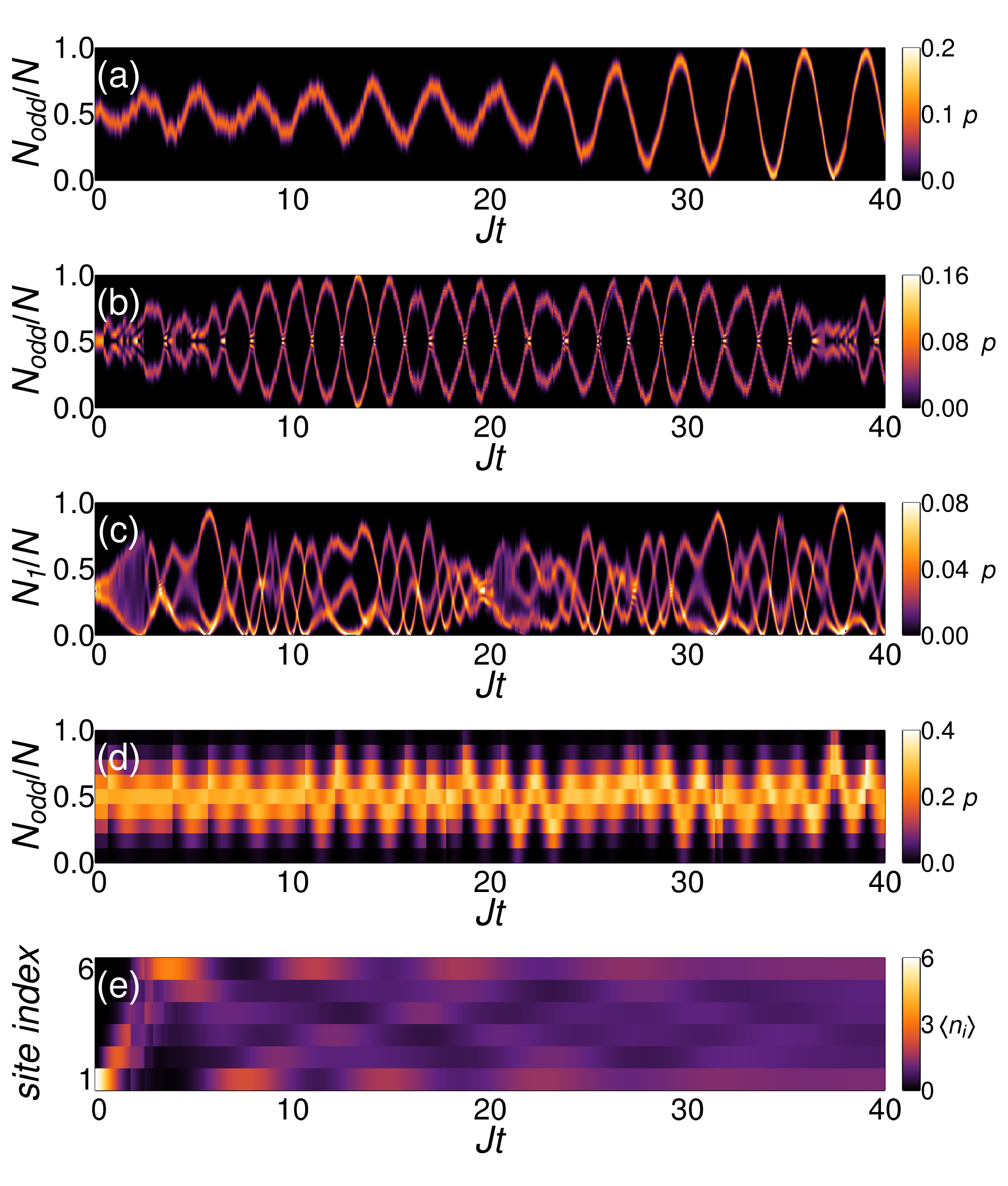}
\end{center}
\caption{Oscilaciones gigantes inducidas en la densidad espacial por medici—n dŽbil. El estado inicial es un estado homogŽneo. Sistemas bos—nicos (a,b,d), sistema fermi—nico (c). (a) Un modo de luz inducido. (b) Dos modos de luz inducidos. (c) Tres modos de luz inducidos. (c) Modo de luz inducido en fermiones.(d) Modo de enlace inducido en bosones. Figura tomada de ~\cite{Mazzucchi1}.}
\label{fig2}
\end{figure}

\subsection{Galer'a de algunos resultados por din‡mica inducida por mediciones }

\begin{figure}[t!]
\begin{center}
\includegraphics[width=0.6\textwidth]{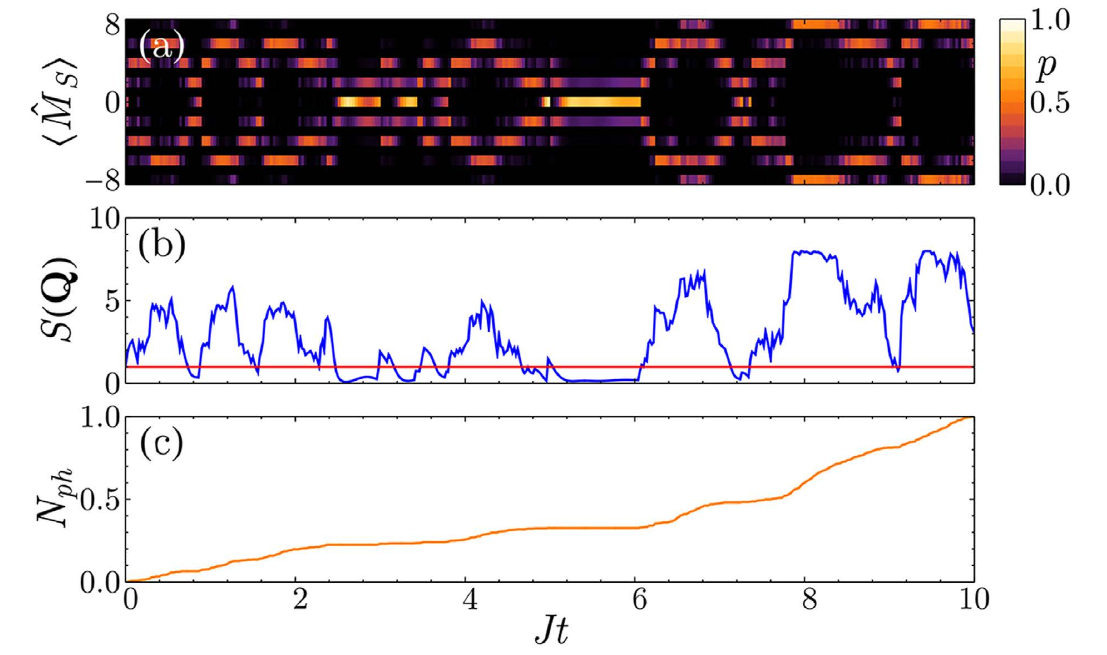}
\end{center}
\caption{Emergencia de orden anti-ferromagnŽtico  en trayectorias con $U=0$ por mediciones dŽbiles en el tiempo. (a) Distribuci—n  de la magnetizaci—n escalonada por sitio en el sistema fermi—nico. (b) Factor de estructura de la magnetizaci—n escalonada. (c) Nœmero de fotones acumulados que escapan del sistema. Figura tomada de ~\cite{Mazzucchi2}.}
\label{fig3}
\end{figure}

En general uno de los efectos de la medici—n dŽbil es inducir el rompimiento de simetr'a espacial din‡mico dependiendo de como los modos de luz inducidos acoplan espacialmente en el sistema ver ~\cite{Mazzucchi, Mazzucchi1}. Se puede inducir diferente nœmero de modos espaciales que en principio son detectables via la medici—n de las poblaciones por sitio~\cite{Mazzucchi1} e incluso otras estructuras con coherencia m‡s complicadas utilizando modos de enlace~\cite{Wojciech2}. Esto es una consecuencia de la estructura del arreglo geomŽtrico de la luz ver ~\cite{Wojciech}. La medici—n compite con la din‡mica t'pica del sistema con procesos de tunelaje y din‡mica at—mica que actœan de manera homogŽnea t'picamente. Con interacciones es posible que el comportamiento de otras fases de la materia cu‡ntica como pueden ser la presencia de aislantes sean modificadas cambiando los valores cr'ticos de la emergencia del superfluido en el sistema de Bose-Hubbard por ejemplo, as' como la destrucci—n o la protecci—n de la existencia de pares en un sistema fermi—nico (Hubbard)  dependiendo del esquema de medici—n, ver ~\cite{Mazzucchi1}. Adicionalmente, es posible estabilizar  —rdenes en el sistema o incluso inducirlos sin interacciones ($U=0$), como pueden ser la emergencia de la magnetizaci—n escalonada en el tiempo en sistemas fermi—nicos con esp'n, Figuras \ref{fig2} y \ref{fig3}, ~\cite{Mazzucchi2}.   En ~\cite{Mazzucchi}, se encuentra que las oscilaciones gigantes inducidas por la medici—n dŽbil son equivalentes a una fuerza estoc‡stica cuasiperi—dica aplicada al sistema. Otra manera de entender el fen—meno se presenta por medio del an‡lisis de estabilidad de modelos efectivos via los exponentes de Lyapunov del sistema estoc‡stico donde es claro que existe un regimen de competencia que a partir de un umbral cr'tico da lugar una inestabilidad que da lugar a que las oscilaciones gigantes en la densidad aparezcan de manera consistente estoc‡sticamente en las trayectorias~\cite{Mazzucchi}. Cabe recalcar que el comportamiento observado a nivel de trayectorias individuales arroja informaci—n que permite establecer que es posible estabilizar dichas oscilaciones. Es interesante que el promedio de las trayectorias arroja resultados triviales, dado que las oscilaciones son de car‡cter aleatorio y tiene fases incoherentes, pero no as' grupos representativos de trayectorias. Efectos de detecci—n ineficiente no eliminan por completo la din‡mica inducida por medici—n para valores de hasta $10\%$ no hay cambios significativos e incluso solo con un $1\%$ de eficiencia se pueden observar lo efectos en las trayectorias para escenarios bos—nicos y fermi—nicos ~\cite{Mazzucchi, Mazzucchi2}. Adicionalmente es posible analizar la estabilidad y controlar la emergencia de estados obscuros por dise–o modificando la din‡mica no-Hermitiana ~\cite{Wojciech1, Wojciech2}. Para calcular los observables del sistema uno puede emplear la siguiente ecuaci—n de movimiento efectiva con trayectorias para un operador $\hat{O}$~\cite{Mazzucchi2},

\begin{figure}[t!]
\begin{center}
\includegraphics[width=0.6\textwidth]{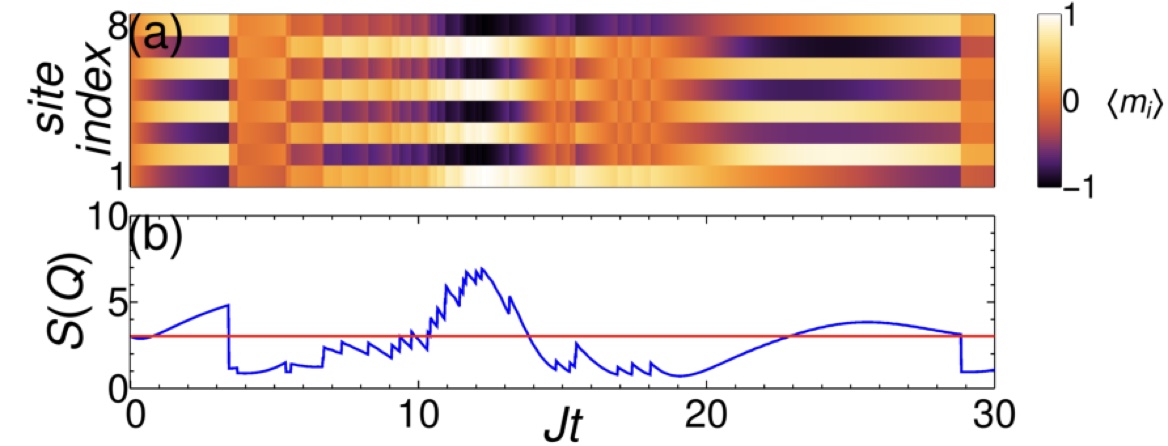}
\end{center}
\caption{Optimizacion de orden anti-ferromagnŽtico con $U>0$ por mediciones dŽbiles en el tiempo. (a) Distribuci—n  de la magnetizaci—n por sitio en el sistema fermi—nico. (b) Factor de estructura de la magnetizaci—n escalonada. Figura tomada de ~\cite{Mazzucchi2}.}
\label{fig4}
\end{figure}

\begin{equation}
\frac{\mathrm{d}}{\mathrm{d}t}\langle\hat{O}\rangle=-i\langle[\hat{\mathcal{H}}^{f/b},\hat{O}]\rangle-\langle\{\hat{c}^\dagger\hat{c},\hat{O}\}\rangle+2\langle\hat{O}\rangle\langle\hat{c}^\dagger\hat{c}\rangle
\end{equation}
para la evoluci—n no Hermitiana y con saltos dados por,
\begin{equation}
\frac{\mathrm{d}}{\mathrm{d}t}\langle\Psi|\Psi\rangle=-\langle\hat{c}^\dagger\hat{c}\rangle
\end{equation}
donde el operador despuŽs del salto cambia tal que,
\begin{equation}
\langle\hat{O}\rangle\to\frac{\langle\hat{c}^\dagger\hat{O}\hat{c}\rangle}{\langle\hat{c}^\dagger\hat{c}\rangle}
\end{equation}

\begin{figure}[t!]
\begin{center}
\includegraphics[width=0.6\textwidth]{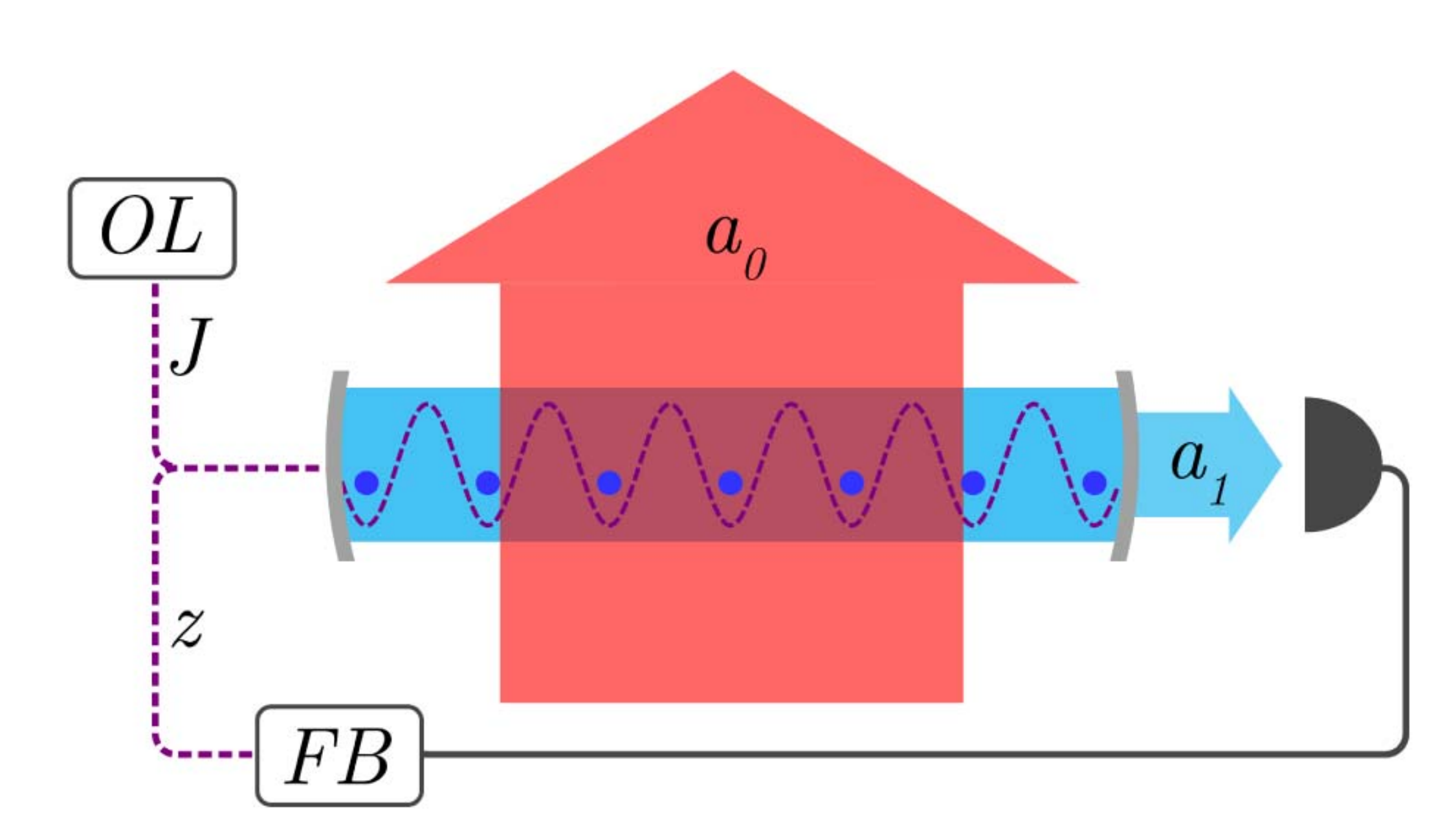}
\end{center}
\caption{Esquema de retroalimentaci—n. Los fotones que escapan la cavidad donde se encuentran ‡tomos en una red —ptica son medidos. La se–al de los fotones que escapan es procesado por elementos electr—nicos y se retro-alimenta el sistema dentro de la cavidad modificando la intensidad de los l‡seres que atrapan los ‡tomos en la red —ptica. Figura tomada de ~\cite{Mazzucchi3}.}
\label{fig5}
\end{figure}

\section{Retroalimentaci—n y control de criticalidad}

M‡s all‡ de solo medir los fotones que escapan el sistema y afectan  el estado cu‡ntico al interior del sistema, se puede considerar la retroalimentaci—n cl‡sica del sistema. Esto se logra procesando la  informaci—n de salida (foto-detecciones) y modificando en tiempo real los par‡metros del sistema de acuerdo a la se–al medida (i.e. la intensidad de los l‡seres que atrapan los ‡tomos en la red —ptica). Un esquema se ilustra en la Figura \ref{fig5}~\cite{Mazzucchi3}. Esto es posible debido a que existe en estos sistemas diferentes rangos de frecuencia de respuesta de cada componente, t'picamente la respuesta electronica sucede en el rango de GHz, la respuesta de la cavidad es en MHz y la respuesta de los procesos at—micos es en kHz. De manera efectiva los cambios realizados para los ‡tomos despuŽs del procesamiento de informaci—n y modificaci—n de par‡metros son instant‡neos. 

\begin{figure}[t!]
\begin{center}
\includegraphics[width=0.6\textwidth]{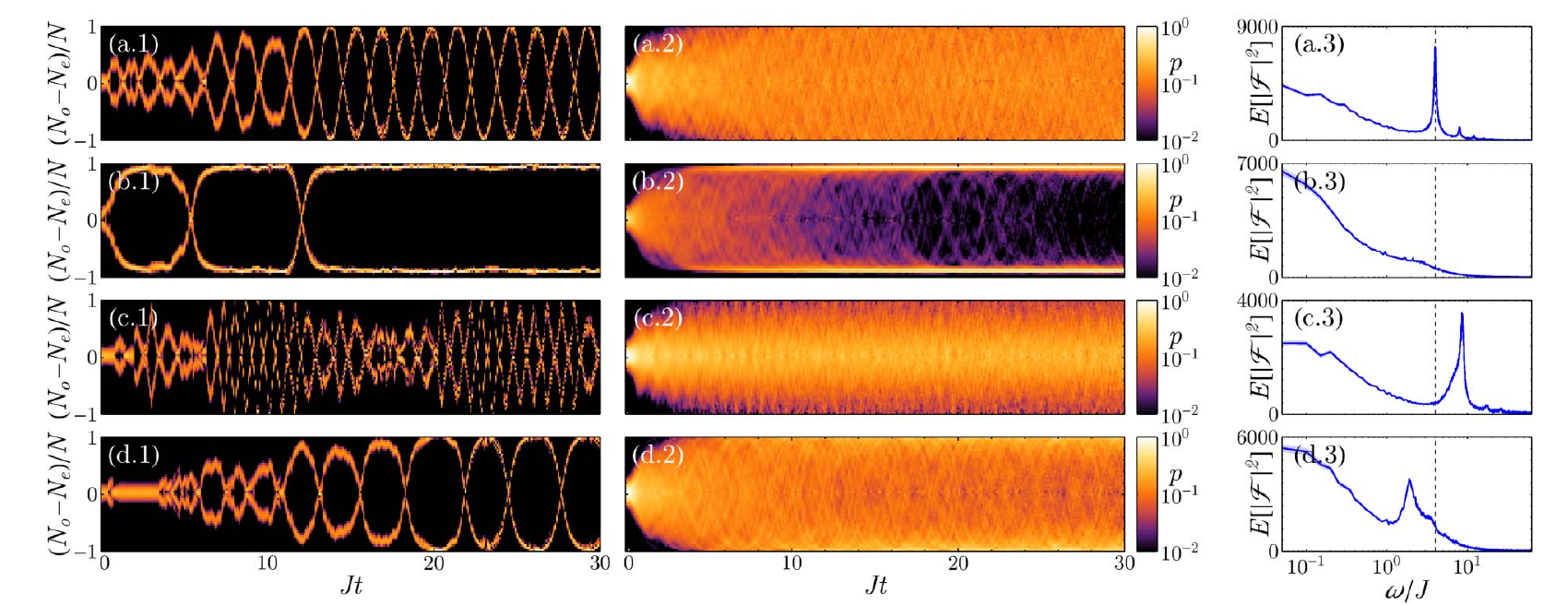}
\end{center}
\caption{Control de la frecuencia de oscilaci—n y congelamiento desbalance arbitrario de poblaciones.(a 1, 2,3 ) Sistema sin retroalimentaci—n. (b,1,2,3) Sistema arriba del l'mite de retroalimentaci—n para congelar el desbalance. (c,1,2,3) Control de la frecuencia de oscilaci—n arriba de la frecuencia natural de oscilaci—n por la din‡mica at—mica.  (d,1,2,3) Control de la frecuencia de oscilaci—n abajo de la frecuencia natural de oscilaci—n por la din‡mica at—mica.   Figura tomada de ~\cite{Mazzucchi3}.}
\label{fig6}
\end{figure}

Utilizando este tipo de sistema es posible en primera instancia estabilizar el desbalance que emerge en el sistema (oscilaciones gigantes de la secci—n anterior) y se puede controlar la frecuencia de las oscilaciones en el sistema inducidas por la pŽrdida de fotones, Figura \ref{fig6}. TambiŽn es posible estabilizar la emergencia de orden anti-ferromagnŽtico en un sistema fermi—nico~\cite{Mazzucchi3}.

Adicionalmente, al considerar la estabilizaci—n y control de estados es posible formular la pregunta si es posible controlar incluso la criticalidad del sistema, modificando exponentes cr'ticos y la posici—n de los mismos. Esto genera un nueva escenario donde se tiene un sistema cu‡ntico que va m‡s all‡ del paradigma disipativo. Recientemente, estudios te—ricos~\cite{Feedback1,Feedback2,Feedback3} han mostrado que es posible. Considerando dos modos en un condensado de Bose-Einstein (BEC) dentro de una cavidad tal que
\begin{equation}
\psi(x)\approx\frac{1}{\sqrt{L}}c_0+\sqrt{\frac{2}{L}}c_1\cos(k_1 x)
\end{equation}
donde $c_0$ corresponde a $k=0$ y $c_1$ a $k=k_1$. DespuŽs de manipulaciones algebraicas se llega al siguiente modelo efectivo~\cite{Feedback1,Feedback2},
\begin{equation}
\mathcal{H}_{\mathrm{eff}}\approx\Delta\hat{a}^\dagger\hat{a}+\omega_R\hat{S}_z+\frac{2}{\sqrt{N}}\hat{S}_x[g(\hat{a}+\hat{a}^\dagger)+GI(t)]
\end{equation}
donde se despreciaron por simplicidad las interacciones de corto alcance entre ‡tomos. Este es de manera efectiva un modelo de Dicke no Markoviano, donde la retroalimentaci—n se considera en $GI(t)$ con 
\begin{eqnarray}
I(t)&\sim&\int_0^th(t-z)\mathcal{F}[x^{\mathrm{out}}_\theta(z)]dz
\\
&=&\int h(t-z)(\hat{a}e^{-i\theta}+\hat{a}^\dagger e^{i\theta})dz
\end{eqnarray}
$I(t)$ es la se–al de control con $G$ el coeficiente de retroalimentaci—n. El kernel $h(x)\sim 1/(t_0+x)^{s+1}$ permite modificar dependiendo de su forma funcional adem‡s del valor del valor cr'tico para la transici—n de fase, la criticalidad del sistema como funci—n de $s$. Se tienen exponentes cr'ticos de valor arbitrario para transici—n de fase que ocurre entre oscilaciones en los valores de expectaci—n del los operadores de materia i.e. $\hat{S}_x\approx\hat{X}$ y un valor estacionario con el ablandecimiento de los modos temporales~\cite{Feedback1}. La criticalidad se extrae a partir de:
\begin{equation}
\lim_{G\to G_c}\langle \hat{X}^2\rangle=\frac{A}{{|1-G/G_\mathrm{crit}|}^\alpha}+B
\end{equation} 
donde $\alpha=\alpha(s)$.

Se ha mostrado que es posible emplear un esquema de retroalimentaci—n que modifica los puntos cr'ticos incluso sin la necesidad de una cavidad en ~\cite{Feedback2}. 

Esquemas similares de retroalimentaci—n han sido  empleados recientemente experimentalmente ~\cite{Esslinger-feedback}. En estos experimentos fue posible estabilizar estados superradiantes supers—lidos con tiempos de vida de m‡s de un orden de magnitud con respecto al sistema sin retroalimentaci—n.

\section{Conclusiones}

Mediante la manipulaci—n en como se miden los fotones que escapan el sistema es posible generar estados cu‡nticos que t'picamente no ocurren. Los estado que emergen de  manera din‡mica presentan propiedades interesantes y rompimientos de simetr'a. Utilizando esta ingenier'a de estados por efectos de medici—n es posible modificar, estabilizar y generar comportamiento colectivo con posibles propiedades œtiles para simulaci—n cu‡ntica. Adicionalmente, el uso de sistemas compuestos con retroalimentaci—n ofrece una flexibilidad aun mayor en tŽrminos de la ingenier'a de propiedades de estos sistemas de materia cu‡ntica. Es posible, adem‡s de preparar estados de manera din‡mica, controlar arbitrariamente como es que se llega a una fase de la materia en particular cerca de puntos cr'ticos, modificar su posici—n y sus leyes de potencia.
La combinaci—n de los sistemas de materia cu‡ntica con la —ptica cu‡ntica da nuevas herramientas para controlar las propiedades fundamentales de estos sistemas con posibles aplicaciones œtiles para la simulaci—n y el control cu‡ntico avanzado.
 
\section{Agradecimientos}
Se agradece a G. Mazzucchi, W. Kozlowski, D. A. Ivanov, T. Ivanova e I. B. Mekhov por contribuciones esenciales en el trabajo descrito en estas notas.

\end{document}